\documentclass[iop,revtex4]{emulateapj}

\newcommand{\nolinkurl}{\url}

\shorttitle{A new equal-area isolatitudinal grid on a spherical surface}
\shortauthors{Z. Malkin}

\begin{document}

\title{A new equal-area isolatitudinal grid on a spherical surface}
\author{Zinovy Malkin$^{1,2}$}
\affiliation{Pulkovo Observatory, St.~Petersburg 196140, Russia}
\affiliation{Kazan Federal University, Kazan 420000, Russia}
\email{zmalkin@zmalkin.com}

\begin{abstract}
A new method SREAG (spherical rectangular equal-area grid) is proposed to divide a spherical surface
into equal-area cells.
The method is based on dividing a sphere into latitudinal rings of near-constant width
with further splitting each ring into equal-area cells.
It is simple in construction and use, and provides more uniform width of the latitudinal rings
than other methods of equal-area pixelization of a spherical surface.
The new method provides a rectangular grid cells with the latitude- and longitude-oriented boundaries,
near-square cells in the equatorial rings, and the closest to uniform width of the latitudinal rings
as compared with other equal-area isolatitudinal grids.
The binned data is easy to visualize and interpret in terms of the longitude-latitude
rectangular coordinate system, natural for astronomy and geodesy.
Grids with arbitrary number of rings and, consequently, wide and theoretically unlimited range of cell size
can be built by the proposed method.
Comparison with other methods used in astronomical research showed the advantages of the new approach
in sense of uniformity of the ring width, a wider range of grid resolution, and simplicity of use.
\end{abstract}

\maketitle


\section{Introduction}

Problem of subdividing (pixelization) of a spherical surface into equal-area cells is quite common in astronomy
and geodesy.
This operation is used to solve different scientific tasks related to data analysis on the sphere, such as
\begin{itemize}
\item averaging data over the cells to mitigate the random errors or to reduce the data dimension;
\item obtaining uniformly distributed data set from unevenly spaced catalogs or measurements;
\item optimizing computation of spherical functions, wavelets and Fourier analysis;
\item comparison and cross-analysis of differently sampled data sets.
\end{itemize}

Many different methods have been developed to construct a regular grid on the celestial sphere or
on the earth's surface.
Detailed description and analysis of these methods can be found, e.g., in
\citet{White1992,Gringorten1992,Gorski2005,Beckers2012,Mahdavi-Amiri2015,Watkins2015}
and papers cited therein. 
These methods use different approaches, may be designed and/or optimized for specific tasks,
and differ in the cell shape, grid geometry, and so on. 
Three methods seem to be most often used in astronomy and geodesy. 

The simplest pixelization of a spherical surface can be realized using the equirectangular projection,
also called equidistant cylindrical projection.
In this grid, all the cells have the same longitude (right ascension) and latitude (declination) span.
The cell area is the same for all the cells belonging to one latitudinal ring, but it is not constant
over the sphere.
It decreases as the distance from the equator increases.

The Lambert azimuthal equal-area projection, also called Lambert zenithal equal-area projection, as well as other
similar methods \citep{Gringorten1992,Rosca2010,Beckers2012} provide subdivision of a plane circular area (disc)
into equal-area cells located on the concentric rings of constant width.
However, being projected on a spherical surface, the grid is distorted and the grid ceases to have
uniform ring width.

The HEALPix pixelization \citep{Gorski2005} is widely used in astronomy during last years.
It subdivides a spherical surface into rhombus-like cells of equal area.
Although the HEALPix grid, unlike other ones discussed in this paper, does not contain latitudinal bands,
one can select chains of cells whose centers have the same latitude, which are considered as latitudinal
rings for the HEALPix grid \citep{Gorski2005}.

Many applications would benefit of using rectangular isolatitudinal grid with the following properties:
\begin{enumerate}
\item it consists of rectangular cells with the boundaries oriented along the latitudinal and longitudinal circles,
which allows to easily connect the binned data with the longitude-latitude coordinate system normally used
in astronomy and geodesy;
\item it has uniform cell area over the sphere;
\item it has uniform width of the latitudinal rings;
\item it has near-square cells in the equatorial rings;
\item it allows simple realization of basic functions such as computation of the cell number given object position,
  and computation of the cell center coordinates given the cell number.
\end{enumerate} 
 
To our knowledge, no such subdivision of a spherical surface has been constructed (and perhaps a rigorous solution 
to this problem does not exist at all).
Strictly speaking, such a solution exists in the form of a sectorial grid with constant longitudinal step,
maybe including the equator.
In the extreme case, there can be only two cells corresponding to the two hemispheres.
However, these solutions are not of practical interest.

A new approach to the solution of the problem of constructing of an optimal grid which best satisfies the goals
listed above was proposed in \citet{Malkin2016a}.
In that paper, three grids were described which were built using an iterative procedure of simultaneous adjustment
of all the latitudinal boundaries, which produced the grids of relatively poor uniformity of the cell area.
Besides it is time consuming for large number of rings, and thus is practically usable only for large-scale grids.

In the current paper, the method proposed in \citet{Malkin2016a} was further advanced.
A new simple but mathematically rigorous method SREAG (spherical rectangular equal-area grid)
was developed which allows us to build a grid of arbitrary
resolution constrained only by the precision of machine calculations.
In Section~\ref{sect:grid_description}, a new strategy is described and discussed in more detail.


\section{A new method of subdivision of a spherical surface}
\label{sect:grid_description}

According to the method discussed in this study, a spherical surface is first split into several latitudinal
rings (bands) of constant width $dB$.
Then each ring is split into several cells of equal size.
The longitudinal span of the cells in each ring is computed as $dL_i = dB \sec b_0^i$,
where $i$ is the ring number, and $b_0^i$ is the central latitude of the ring.
This provides near-square cells in the equatorial rings.
Then the number of cells in each ring equal to $360 /dL_i$ is rounded to the nearest integer value.
This procedure results in the initial grid.

The initial grid satisfies the requirements 1 and 3--5 given above, but does not satisfy item 2.
However, the latter requirement is considered as critical in most practical applications.
Although there is no known pixelization method that would provide equal cell area and equal width of the latitudinal
rings simultaneously, an approximate solution to this problem can be proposed given that the the uniformity
of the cell area is the top priority.
To achieve this goal, the latitudinal boundaries of the rings should be adjusted to provide uniform cell area over
the sphere.
Notice that the cell span in the longitudinal direction remains fixed to the values computed for the initial grid.

The adjustment procedure is the following.
Given the number of rings $N_{ring}$.
Let $A$ be the cell area computed as the area of the unit sphere $4\pi$ divided by the number of cells in the grid $N_{cell}$. 
Let us start from the North pole.
Let $b^u$ be the upper (closer to the pole) boundary of the ring in the final (adjusted) grid, and $b^l$ be
the lower boundary.
Then, taking into account that the cell area is $A = dL ( \sin b^u - \sin b^l )$, the simple loop will
allow to compute all the ring boundaries:

\indent $b^u_1 = \pi/2$\\         
\indent do i=1,$N_{ring}/2$\\
\indent\indent $b^l_i = \arcsin( \sin b^u_i - A/dL_i )$\\
\indent\indent $b^u_{i+1} = b^l_i$\\
\indent end do\\
         
The last value $b^l_{Nring/2}$ must be equal to zero (corresponds to the equator), which verifies the correctness
of the computation.
After that, the latitudinal boundaries for the rings in the South hemisphere are just copied from the North
hemisphere with opposite (negative) sign.

\begin{figure}
\centering
\includegraphics[clip,width=\hsize]{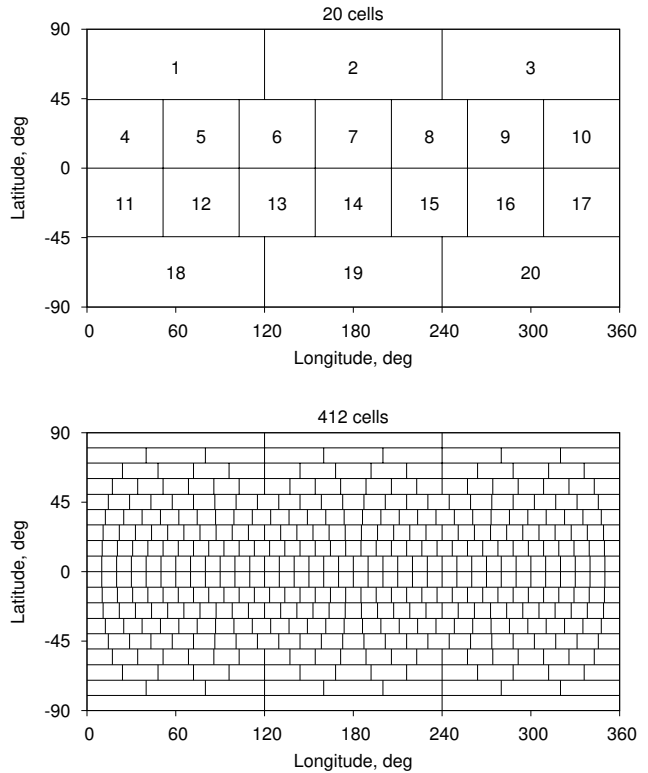}
\caption{Two examples of the grids obtained using the SREAG method.
  The upper grid consists of 4 rings and 20 cells, the lower grid consists of 18 rings and 412 cells.
  The cell numbering scheme is shown in the first grid.}
\label{fig:cells}
\end{figure}

Figure~\ref{fig:cells} presents two examples of grids constructed making use of the proposed method.
These grids consist of 20 and 412 cells corresponding to grids with 4 and 18 rings, respectively.

The basic parameters of the first 24 grids are given in Table~\ref{tab:grids}.
Those are the number of rings, the number of cells, cell area, and the maximum difference
between the actual and nominal central latitude $b-b_0$ among all the rings in the given grid.
The nominal central latitude of the ring is the central ring latitude for the grid with the same number of rings
of constant width.
In other words, the nominal central latitude is the central latitude of the rings in the initial grid (see above). 

\begin{table}
\centering
\caption{Basic parameters of the first 24 grids: the number of rings ($N_{ring}$), the number of cells ($N_{cell}$),
  cell area ($A$), sq. deg, and the maximum difference between the actual and nominal central latitude ($b-b_0$), deg.}
\label{tab:grids}
\begin{tabular}{rrrc}
\hline
\hline
$N_{ring}$ & $N_{cell}$ & \multicolumn{1}{c}{$A$} & max($b-b_0$) \\ 
\hline
 4 &   20 & 2062.6 & 0.29 \\
 6 &   46 &  896.8 & 0.93 \\
 8 &   82 &  503.1 & 0.24 \\
10 &  128 &  322.3 & 0.38 \\
12 &  184 &  224.2 & 0.37 \\
14 &  250 &  165.0 & 0.34 \\
16 &  326 &  126.5 & 0.34 \\
18 &  412 &  100.1 & 0.34 \\
20 &  508 &   81.2 & 0.33 \\
22 &  614 &   67.2 & 0.31 \\
24 &  732 &   56.4 & 0.34 \\
26 &  860 &   48.0 & 0.36 \\
28 &  998 &   41.3 & 0.20 \\
30 & 1146 &   36.0 & 0.19 \\
32 & 1302 &   31.7 & 0.18 \\
34 & 1466 &   28.1 & 0.16 \\
36 & 1654 &   24.9 & 0.17 \\
38 & 1838 &   22.4 & 0.16 \\
40 & 2038 &   20.2 & 0.15 \\
42 & 2248 &   18.4 & 0.15 \\
44 & 2464 &   16.7 & 0.14 \\
46 & 2696 &   15.3 & 0.13 \\
48 & 2938 &   14.0 & 0.13 \\
50 & 3186 &   12.9 & 0.12 \\
\hline
\end{tabular}
\end{table}

For comparison, the latitudinal distribution of the cells for two other widely used methods is presented
in Table~\ref{tab:central_latitude}. 
These methods include the Lambert equal-area projection (\citet{Gringorten1992,Rosca2010}; 
data in Table~\ref{tab:central_latitude} are taken from the former paper) and two HEALPix grids \citep{Gorski2005}. 
The number of latitudinal rings is taken to be close to one of the grids in Table~\ref{tab:grids}. 
Software provided by the HEALPix team\footnote{http://healpix.sourceforge.net} was used in the computations
related to the HEALPix grids. 
Notice that the subdivision of \citet{Gringorten1992} based on the Lambert projection includes polar caps,
and the HEALPix grids include a single equatorial ring. 
Table~\ref{tab:central_latitude} presents a comparison of the actual central latitude of the rings for two types of
grids with nominal central latitude of the rings.
The difference between the actual and nominal central latitudes reflects the deviation of the width of the rings
for the given grid from the grid with uniform ring width.
Comparison of the results presented in Tables \ref{tab:grids} and \ref{tab:central_latitude} shows that the deviation
of the central latitude of the rings from the uniform distribution is much smaller for the new pixelization method
than for the HEALPix and Lambert methods.
In other words, the proposed method provides a much better approach to the grid with uniform ring width.

\begin{table*}
\centering
\caption{Actual and nominal (corresponding to the grid with uniform ring width) central latitude of the latitudinal
  rings for different subdivision of a spherical surface with the number of cells close to one of the new grids.
  Only the North hemisphere is shown, the South hemisphere is symmetric with respect to the equator.}
\label{tab:central_latitude}
\begin{tabular}{lrrrrrrr}
\hline
\hline
&\multicolumn{7}{c}{Latitudinal rings} \\
\hline
\multicolumn{8}{c}{Lambert projection, 121 cells} \\
Actual  & 75.20   & 60.13  & 44.52  & 27.91  & 9.65 && \\                     
Nominal & 73.64   & 57.27  & 40.91  & 24.55  & 8.18 && \\                           
Diff.   &  1.56   &  2.86  &  3.61  &  3.36  & 1.47 && \\
\multicolumn{8}{c}{HEALPix, $N_{side}$ = 2, 48 cells} \\
Actual  & 66.44   & 41.81  & 19.47  &&&& \\    
Nominal & 77.14   & 51.43  & 25.71  &&&& \\          
Diff.   & --10.70 & --9.62 & --6.24 &&&& \\
\multicolumn{8}{c}{HEALPix, $N_{side}$ = 4, 192 cells} \\
Actual  & 78.28   & 66.44  & 54.34  & 41.81  & 30.00  & 19.47  & 9.59   \\
Nominal & 84.00   & 72.00  & 60.00  & 48.00  & 36.00  & 24.00  & 12.00  \\   
Diff.   & --5.72  & --5.56 & --5.66 & --6.19 & --6.00 & --4.53 & --2.41 \\
\hline
\end{tabular}
\end{table*}

\begin{figure}
\centering
\includegraphics[clip,width=\hsize]{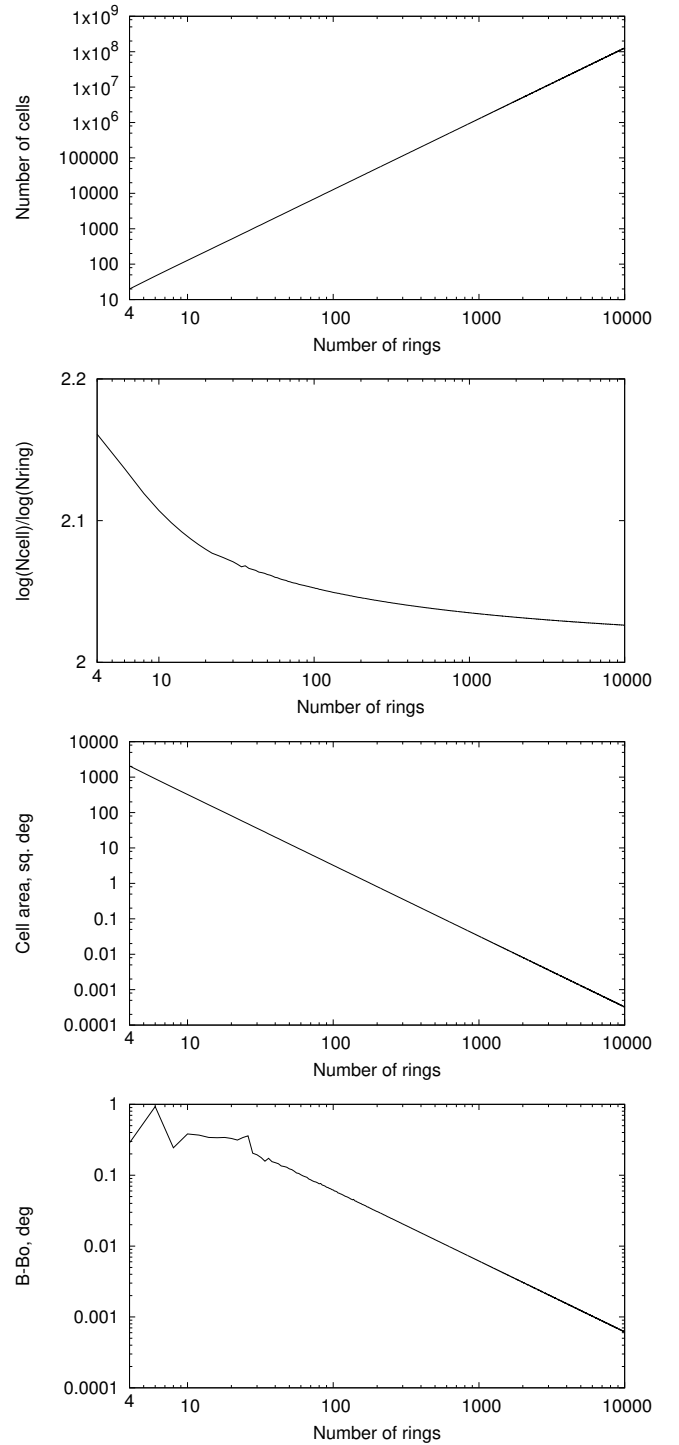}
\caption{Basic parameters of the grids of different resolution (from the top): the number of cells as function
  of the number of rings, $\log(N_{cell})/ \log(N_{ring})$, cell area, the deviation of the actual central
  latitude of the rings from the uniform distribution.}
\label{fig:grids3}
\end{figure}

The basic parameters of the grids constructed using the SREAG method for the whole reasonable range of
the grid resolution are shown in Fig.~\ref{fig:grids3}.
Notice that the proposed method provides much more detailed choice of the grid resolution than,
e.g., widely used HEALPix approach.
The first panel of the plot shows nearly linear log-log relationship between the number of rings
and the number of cells.
Actually, the factor $\log(N_{cell})/ \log(N_{ring})$ varies slightly as shown in the second panel
of Fig.~\ref{fig:grids3}.
Two other panels of the plot show the cell area and the deviation of the actual central latitude
of the rings from the uniform distribution.
The latter show a near monotonic decrease with increasing of $N_{ring}$ for the grids with $N_{ring} > 24$.

\section{Conclusion}
\label{sect:conclusion}

In this paper, a new method, SREAG, is developed for subdividing a spherical surface into equal-area isolatitudinal
cells with near-uniform latitudinal distribution. 
Advantages of the proposed approach to construction of the uniform grid on the sphere are
\begin{itemize}
\item it provides a strictly uniform cell area;
\item it provides a rectangular grid cells with the latitude- and longitude-oriented boundaries with
  near-square cells in the equatorial rings;
\item it provides a wide, theoretically unlimited  range of grid resolution with much more detailed choice
  of desirable cell area than widely used HEALPix method;
\item it is simple in realization and use
 (Fortran routines are available\footnote{http://www.gaoran.ru/english/as/ac\_vlbi/index.htm\#SREAG});
\item the binned data is easy to visualize and interpret in terms of the longitude-latitude
  (right ascension-declinations) rectangular coordinate system, natural for astronomy and geodesy.
\end{itemize}

The gird consists of latitudinal rings divided into uniform rectangular cells with the boundaries directed
along meridians, which makes it easy to define the pixel that contains the given point
and solve the inverse problem of computation of the coordinates of the cell from the cell number. 

Although the grids of the same type described in \citet{Malkin2016a} have low accuracy as compared with the grids
built using the rigorous method derived in the present work, they were used in several applications to date.
The division of the sky into 46 cells was used in \citet{Malkin2016b} to re-sample a non-uniform distribution
of radio stars over the sky into uniformly distributed sample.
The grid of 130 cells (similar to the more accurate 128-cell grid mentioned above) occurred to be convenient
to analyze the distribution of giant radio sources over the sky \citep{Kuzmicz2018}.
The 406-cell grid proposed in \citet{Malkin2016a} (similar to the more accurate 412-cell grid mentioned above)
provides a subdivision of the sphere into the cells of the area approximately equal to 10~deg$^2$.
It was found to be appropriate to investigate the distribution of the sources in the third realization of
the International Celestial Reference Frame (ICRF3, \citet{Jacobs2014}) sources over the sky \citep{Basu2019}.

To our knowledge, the pixelization method described in this paper provides the best compromise between
uniform cell area and uniform cell latitudinal width.
Priority is given to equal cell area, as it looks more important for practical applications.
Other grids with central ring and/or polar caps can be also constructed using the method proposed in this study.

The proposed SREAG approach to pixelization of a celestial or terrestrial spherical surface allows us to construct
a wide range of grids for analysis of both large-scale and tiny-scale structure of data given on a sphere. 
The number of cells is theoretically unlimited and is constrained in practice only by the precision of machine
calculations.
Such unique properties hopefully make the method useful for various practical applications in different research fields
in astronomy, geodesy, geophysics, geoinformatics, and numerical simulation. 
In particular, it can be used in further analyses of the celestial reference frame, for selection of uniformly
distributed reference sources in the next ICRF realizations, and for evaluation of the systematic errors of the source
position catalogs.


\section{Acknowledgments}

The author thanks the anonymous reviewer, the editor, Dr. Chris Lintott, and Anton Malkin for valuable comments
and suggestions which helped to improve the manuscript.

This work was partly supported by the Russian Government program of Competitive Growth of Kazan Federal University.

This research has made use of NASA's Astrophysics Data System.

\bibliography{sky_cell}
\bibliographystyle{aasjournal}

\end{document}